\documentclass[11pt]{article}

\usepackage[final]{acl}

\usepackage{times}
\usepackage{latexsym}

\usepackage[T1]{fontenc}

\usepackage[utf8]{inputenc}

\usepackage{microtype}

\usepackage{inconsolata}

\usepackage{graphicx}

\usepackage{booktabs}       
\usepackage{xcolor}         
\usepackage{listings}       
\colorlet{punct}{red!60!black}
\definecolor{background}{HTML}{EEEEEE}
\definecolor{delim}{RGB}{20,105,176}
\colorlet{numb}{magenta!60!black}
\lstdefinelanguage{json}{
    basicstyle=\normalfont\ttfamily,
    showstringspaces=false,
    breaklines=true,
    frame=lines,
    string=[s]{"}{"},
    stringstyle=\color{blue},
    comment=[l]{:},
    commentstyle=\color{black},
}
\usepackage{makecell}  
\usepackage{subcaption}  
\usepackage{graphicx}  
\usepackage[
]{caption}
\title{SchemaRAG: Dynamic Large Schema Reduction for LLM-driven Structured Information Extraction}

\author{
  \textbf{Sin Yu Bonnie Ho}\thanks{Equal contribution. Corresponding authors: \texttt{siho@microsoft.com}, \texttt{arliecoles@microsoft.com}} \\ \And
  \textbf{Arlie Coles}\footnotemark[1] \\ \And
  \textbf{Erik Larsson} \\ \AND
  \textbf{Eric Marshall} \\ \And
  \textbf{Nathan Bodenstab} \\ \And
  \textbf{Paul Vozila} \\ \AND
  \textnormal{Microsoft}
}

\begin{document}
\maketitle
\begin{abstract}
Extracting structured data from unstructured text using large language models (LLMs) becomes challenging when the target schemas are large and complex. 
In such cases, including the full schema in the prompt 
increases cost and latency, risks lost-in-the-middle performance degradation, and can exceed context length limits. 
We propose SchemaRAG, a retrieval-augmented generation (RAG) framework that dynamically prunes the output schema space for schema-conditioned information extraction tasks by leveraging schema metadata and few-shot examples (when available).
We evaluate SchemaRAG on real-world healthcare and e-commerce datasets. 
Our results show that SchemaRAG can achieve up to an 8.8\% increase in micro-F\textsubscript{1}, 
a 47\% reduction in latency, and a 48\% reduction in token costs,
demonstrating its practicality for large-schema extraction.
\end{abstract}

\section{Introduction}

Structured information extraction (IE) pairs values from unstructured text with schema-defined keys. In healthcare, 
this includes medical attribute extraction from unstructured clinical notes \cite{Jiang2011-qt, agrawal-etal-2022-large} and accelerated human annotation of medical narrative data \cite{pmlr-v225-goel23a}. 
Similarly, e-commerce applications focus on product attribute value extraction from unstructured text descriptions with and without supplementary multimodal data \cite{brinkmann2024,zhu-etal-2020-multimodal}.

Further applications of structured IE exist for problems that can be cast as form-filling, 
where models need to populate complex, hierarchical schemas from unstructured text. 
These forms could have numerous rows, and each row could have complex constraints on potential values (\textit{e.g.}, required types including strings, numbers, single-select radio buttons, multi-select checkboxes, or other custom types). In this work, we consider two such use cases: 
1) populating electronic health records (EHRs) from medical narratives and 2) filling a product metadata form from e-commerce product description text.

Instruction-tuned large language models (LLMs) are well-suited for these tasks due to their broad semantic knowledge \cite{Singhal2023} and ability to follow structured prompt instructions \cite{LIU2024103809}.
However, naively injecting large schemas into the prompt can exceed context limits or introduce many irrelevant rows. This leads to lost-in-the-middle effects \cite{liu-etal-2024-lost} and performance degradation as the model struggles to isolate pertinent rows. The resulting latency and inference costs can preclude real-time applications.

\begin{figure*}
    \centering
    \includegraphics[width=\textwidth]{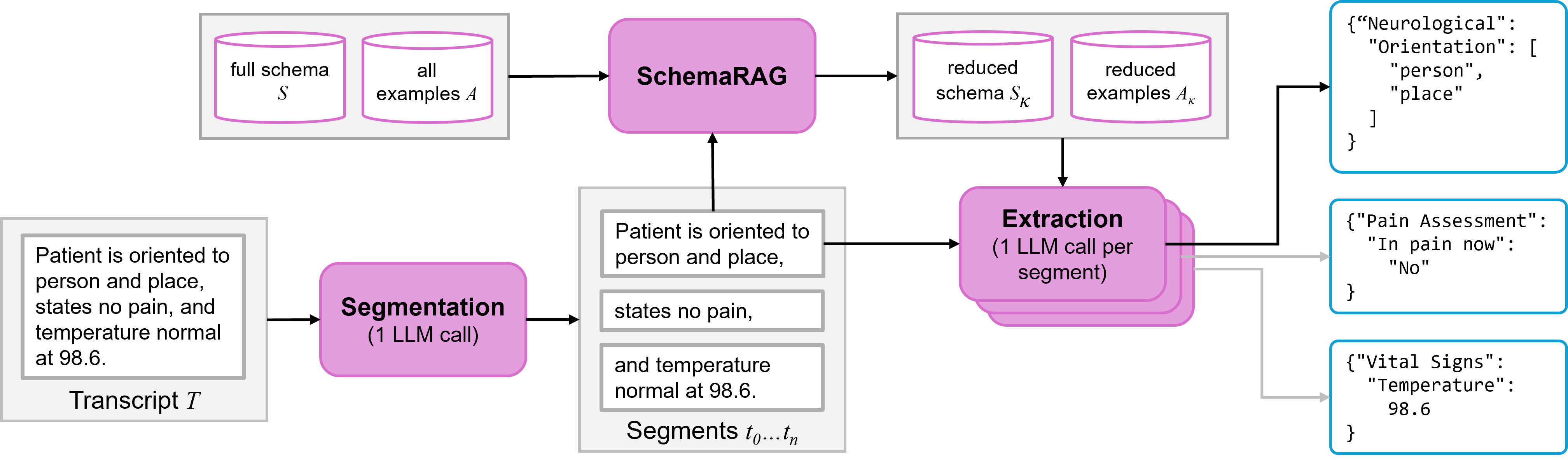}
    \caption{Overview of SchemaRAG which extracts information by segmenting unstructured text and applying schema reduction to each segment. These segments, alongside their reduced schemas and any retrieved examples, are then passed to a just-in-time prompted LLM query for extraction.}
    \label{fig:overview}
\end{figure*}

To mitigate these issues, we propose Schema-RAG, a retrieval-augmented generation approach that retrieves and prunes the output schema space. By leveraging schema metadata and, when available, limited labeled data, SchemaRAG dynamically selects a relevant schema subset for just-in-time in-context prompted LLM use (Figure \ref{fig:overview}). 
A well-selected reduced schema shrinks prompt size, decreases the chance of the LLM filling in irrelevant rows, and improves overall extraction performance. 

The framework is schema-agnostic and supports arbitrary levels of hierarchy. Because the embeddings driving the retrieval can be computed offline, the system is efficient and easily integrated into existing pipelines. Crucially, SchemaRAG operates as a training-free solution, requiring no pretraining or fine-tuning. 
We explicitly focus on schema-conditioned extraction using hosted LLMs under real-world deployment 
scenarios
including large, complex schemas and strict resource budgets. 
By utilizing prompting rather than retraining, our approach allows for rapid iteration and immediate adaptability to evolving requirements.

The contributions of the paper are as follows:
1) We propose SchemaRAG---a RAG-based method to reduce the solution space for LLM-driven structured information extraction tasks with large, complex output schemas.
2) We publish experimental results showing that our method delivers  up to an 8.8\% relative micro-F\textsubscript{1} improvement on two real-world, large-schema datasets. Latency and token cost reduce by up to 47\% and 48\% respectively, though outcomes vary by dataset.

\section{Related Work}

LLMs have been widely used for IE tasks including named entity recognition (NER), relation extraction (RE), and attribute value extraction (AVE) \cite{Dagdelen2024, wang2023instructuiemultitaskinstructiontuning, wei2024chatiezeroshotinformationextraction, brinkmann2024, ZOU2025144572,wu-etal-2024-learning}. 
Recent retrieval-based enhancements further improve IE by augmenting prompts with relevant context. 
Retrieval-augmented generation (RAG) \cite{lewis2021retrievalaugmentedgenerationknowledgeintensivenlp} refers to techniques that augment LLM prompts with information retrieved from external knowledge sources. A common use case is the dynamic selection of relevant text-based document chunks or few-shot examples for conditioning \cite{
ma2025knowledgegraphbasedretrievalaugmentedgeneration, 10.1007/978-3-031-72344-5_17}.

A parallel line of research retrieves already populated data structures (\textit{e.g.}, tables or databases) and injects subsets into the prompt of a downstream task \cite{chen2024tablerag, sui-etal-2024-tap4llm, lin-etal-2023-inner, ye2023large, wang2024chain}. 
Prior table-centric methods assume available cell values and operate over filled tables to answer questions or perform reasoning. 
Unlike table‑centric QA over populated tables, our setting is schema‑conditioned form‑filling with a large, unpopulated schema to be populated from free text.
Furthermore, we argue that traditional IE benchmarks are unsuitable for evaluating SchemaRAG because their rigid 
span-based
scoring penalizes semantically correct LLM extractions that happen to differ from the dataset's specific tagging biases.
 
These challenges highlight an underexplored area in existing RAG literature: current retrieval strategies do not address the challenge of dynamically selecting relevant portions of a complex, structured output schema to guide LLM-based extraction. Our SchemaRAG method applies retrieval to the output schema itself. SchemaRAG narrows the prompt to a relevant subset of a large, hierarchical schema by retrieving: i) a subset of candidate schema rows based on interpolated row metadata and ii) relevant schema-row-labeled text examples (when available). 
The retrieved rows compose the schema subset injected into the prompt; associated labeled
data (if any) serve
as few-shot examples.

\section{Methodology}

To perform structured data extraction of key-value pairs on a transcript $T$ of unstructured text, extracted pairs must conform to some schema $S$. We define $S$ as composed of $m$ rows, where each row $r_i$ has an identifier $i$, a string name $n_i$, and a subschema describing its possible values $v_i$. The value subschemas of $S$ can be arbitrary, according to the types required by the domain (Table \ref{tab:types}). 

In a complex schema, there may exist arbitrary levels of hierarchy that form a path to any row. 
In this work we consider 2-level schemas, where each row may also belong to any number of categories $C_i$, with each category having a string name $c_j$. This structure is common in EHR systems, but our method is generalizable to schemas with any number of hierarchical levels.

\begin{table}
    \centering
    \caption{Example value subschemas for different row types. 
    }
    \label{tab:types}
    \begin{tabular}{ll}
        \toprule
        \textbf{Type} 
            & \textbf{Value subschema ($v_i$)} \\
        \midrule
        \makecell[lt]{Single}
            & \makecell[lt]{
                \small\texttt{\{"type": "string",} \\ 
                \small\texttt{ "enum": ["Yes", "No"]\}}
            } \\
        \makecell[lt]{Multi}
            & \makecell[lt]{
                \small\texttt{\{"type": "array",} \\
                \small\texttt{ "enum": ["A", "B", "C"]\}}
            } \\
        Numeric 
            & \small\texttt{\{"type": "number"\}} \\
        \bottomrule
    \end{tabular}
\end{table}

$S$ can be represented in string form as a JSON Schema (Table \ref{tab:exampleschema}). 
In the simplest case, $S$ and $T$ can be used directly as in-context input to a prompted LLM query to perform the extraction.
We only use the JSON Schema representation in the prompt,
since LLMs may more readily fill this standard specification format than others \cite{brinkmann2024}.
When annotated examples are available, they can be included in the prompt as in-context learning material. 
In this work, we aim to reduce $S$ to $S_\kappa$, a smaller set of $\kappa \ll m$ most relevant rows, 
in order to both improve performance on the structured extraction task and avoid the problems associated with using a very large $S$ in context.

\subsection{Schema reduction}

Our schema reduction method 
leverages two sources of row information: the metadata of the rows themselves, and, when available, example transcripts annotated by row. For each row $r_i \in S$, we create a row embedding ${e_\mathrm{row}}_i$ by normalized weighted interpolation of embedded row metadata $M_i$, using a pretrained embedder $H$ (Eq. \ref{eq:rowembedding}). 
Intuitively, this is a weighted average over the embeddings of each available metadata field for the row, where the weights control the relative importance of different metadata types.
The contents of $M_i$ are configurable based on available metadata, but minimally include the row name $n_i$, a string concatenation of the row's category names, $C'_i$, and (for Single or Multi rows) a string concatenation of possible values from the row's value subschema, $v'_i$. We treat each weight $w_j$ as a hyperparameter to be tuned by random or other search.

\begin{equation}
    \small
    {e_\mathrm{row}}_i = \sum_{j=1}^{|M_i|} w_j \cdot H({{m_i}_j}) \quad \forall {{m_i}_j} \in M_i = \{n_i, C'_i, v'_i, ...\}
    \label{eq:rowembedding}
\end{equation}

When a set of annotated examples $A$ is available, we also create an example embedding ${e_\mathrm{ex}}_i$ for each example $a_i = (x_i, y_i) \in A$, where $x_i$ is the unstructured text input and $y_i$ is the annotated structured output. This example embedding is simply the embedding of the text input, $H(x_i)$. Both row and example embeddings are within the same vector space, enabling direct similarity comparison.

We compute a relevance score $\rho_i$ of each embedding $e_i \in E = E_\mathrm{row} \cup E_\mathrm{ex}$ as its cosine similarity to the embedding of the transcript, $H(T)$. We build the reduced schema $S_\kappa$ by examining the embeddings sorted by descending $\rho$. For each embedding $e_i$, if it is a row embedding, we add the corresponding row to $S_\kappa$. If it is an example embedding, we add the corresponding example's annotated rows from $y_i$ to $S_\kappa$. When $k$ embeddings have been examined, we stop. 
This $k$-nearest neighbors search in the embedding space may lead to more or fewer than $k$ rows in $S_\kappa$, since multiple embeddings may map to the same row and example annotations may include multiple rows.

\begin{figure*}
    \centering
    \includegraphics[width=\textwidth]{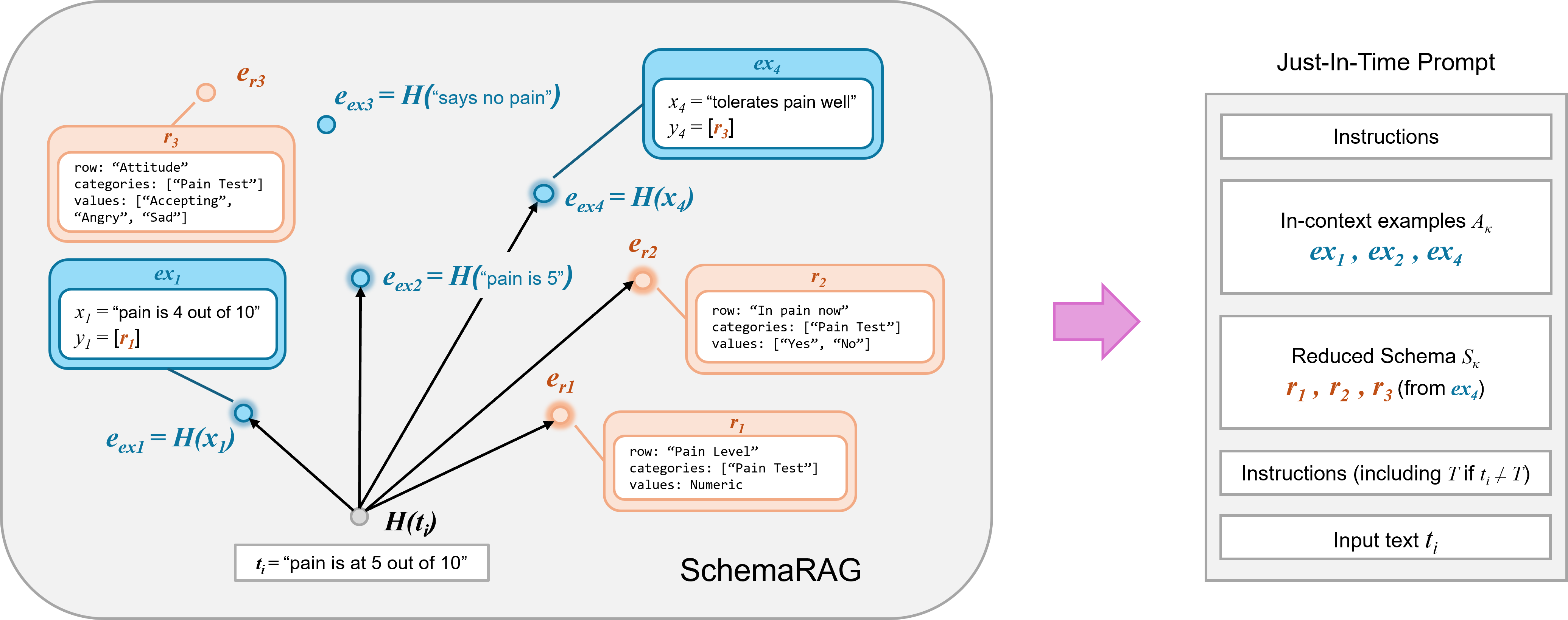}
    \caption{Schema reduction by SchemaRAG (with $k =$ 5). The five closest embeddings to $H(t_i)$ are retrieved. Rows whose embeddings are found are added directly to the reduced schema, and examples whose embeddings are found have their annotated rows added to the reduced schema. The reduced schema is then converted to JSON Schema format for use in the extraction prompt, and the found examples are added as in-context learning material.}
    \label{fig:reduction}
\end{figure*}

This process, visualized in Figure~\ref{fig:reduction}, yields a reduced schema $S_\kappa \subseteq  S$ that is sorted by relevance to the transcript $T$ and can be converted to JSON Schema format for use in-context when performing the extraction task. An example of $S_\kappa $ and its JSON Schema representation as presented to the LLM given the transcript $T$ is provided in Appendix \ref{sec:schema_representation}.

\begin{table}
        \centering
        \caption{Example schema $S$ containing one Single row $r_i$, in JSON Schema format.}
        \label{tab:exampleschema}
        \small
\begin{lstlisting}[language=json,firstnumber=1,columns=fullflexible]
{"type": "object",
 "properties": {
   "category (c_j)": {
      "type": "object",
      "properties": {
        "row (n_i)": {
          "type": "string",
          "enum": ["A", "B", "C"]
}}}}}
\end{lstlisting}
\end{table}

\subsection{Transcript segmentation}

In domains with long unstructured text transcripts, LLMs may struggle to locate all extractable content across extensive text. If the SchemaRAG hyperparameter $k$ is too large, the resulting schemas may cause lost-in-the-middle effects, negating the benefits of schema reduction. Conversely, a $k$ that is too small may fail to cover all relevant information.
To counteract these issues, we experiment with segmenting the transcript into $n$ segments, $t_1, t_2, ..., t_n$, then run SchemaRAG on each segment to obtain per-segment reduced schemas $S_{\kappa_1}, S_{\kappa_2}, ..., S_{\kappa_n}$, and finally run $n$ extractions, with each segment $t_i$ and its corresponding reduced schema $S_{\kappa_i}$ injected into the prompt of each extraction LLM query. Segmentation is performed by a prompted LLM call containing the transcript $T$ and instructions describing how to segment.

\section{Experiments}

\subsection{Datasets}

\textbf{Nursing.}~ We evaluate our method on four proprietary test sets from Hospitals A--D (together the Nursing dataset), each containing between 48--50 transcripts of human-provided, de-identified nurse rounding dictations. Each transcript is human-annotated with a set of key-value pairs from per-hospital schemas $S_{A}$--$S_{D}$ respectively. 
No row in any schema is present in any other schema. All four hospital schemas use ten row types, including Single, Multi, Numeric, and other proprietary custom types. 

For each hospital, we also curate example sets, $A_{A}$--$A_{D}$, linking key-value pairs with their associated transcript provenance. 
We examine the effect of these example sets on our schema reduction method and on the downstream extraction task as in-context learning material. The breakdown of the size and content of Nursing can be found in Appendix \ref{sec:datasets}.

\textbf{Amazon.}~ For variety in domain, we also evaluate our method on a subset of the publicly available sample of the Bright Data Amazon Products dataset \cite{brightdata2025}. 
From 1000 scraped Amazon product webpages, we sample 48 products and 
create a structured extraction task by assembling a transcript for each product through concatenation of the title, description, top review, and key prose attributes. We construct schema rows and annotations through this process and 
manually verify that annotated key–value pairs are supported by the transcript. The resulting test set contains 804 annotated row–value pairs (554 unique rows). 
From the full 1000-product sample, we assemble a schema $S_{\mathrm{Am}}$ with 1906 rows spanning four row types (Single, Multi, Numeric, Weight). All annotated rows in the test set appear in this schema. Amazon transcripts are roughly 10$\times$ longer than Nursing transcripts. 
Full details of the test set and schema construction process are provided in Appendix~\ref{sec:datasets}.

\begin{table*}[t]
\centering
\caption{Overall extraction accuracy. Statistical significance of macro-F\textsubscript{1} improvements over the full schema baseline is indicated by * ($p <$ 0.05) and \textsuperscript{\textdagger} ($p <$ 0.001).}

\label{tab:results}
\begin{tabular}{lrrrr}
    \toprule
    & \multicolumn{2}{c}{\textbf{Nursing}} & \multicolumn{2}{c}{\textbf{Amazon}}  \\
    \cmidrule(lr){2-3}
    \cmidrule(lr){4-5}
    & micro-F\textsubscript{1} $\pm \textsc{cv}$ 
    & macro-F\textsubscript{1} $\pm \textsc{cv}$ 
    & micro-F\textsubscript{1} $\pm \textsc{cv}$ 
    & macro-F\textsubscript{1} $\pm \textsc{cv}$ \\
    \midrule
    Full schema 
        & 0.844 $\pm$ 0.29\%
        & 0.851 $\pm$ 24.7\%
        & 0.471 $\pm$ 3.12\%
        & 0.481 $\pm$ 47.8\% \\
    \hspace{0.4em} + segmentation 
        & 0.880 $\pm$ 0.54\%
        & 0.856* $\pm$ 24.1\% 
        & 0.423 $\pm$ 1.23\%
        & 0.451 $\pm$ 32.4\%\\ 
    \midrule
    SchemaRAG & 
        \textbf{0.918} $\pm$ 0.22\%
        & \textbf{0.899}\textsuperscript{\textdagger} $\pm$ 22.7\% 
        & \textbf{0.510} $\pm$ 0.87\%
        & \textbf{0.515} $\pm$ 34.4\% \\
    \hspace{0.4em} \textit{+ oracle} 
        & \textit{0.952  $\pm$ \textit{0.05}\%} 
        & \textit{0.951\textsuperscript{\textdagger} $\pm$ 13.9\%} 
        & \textit{0.775} $\pm$ \textit{0.35}\%
        & \textit{0.788}\textsuperscript{\textdagger} $\pm$ 14.8\% \\
    \bottomrule
\end{tabular}
\end{table*}

\subsection{Models and hyperparameters}
\label{subsec:models_and_hyperparameters}



For all the experiments, we use OpenAI's \texttt{text-embedding-ada-002} \cite{openaiImprovedEmbedding} as the embedder $H$ and \texttt{gpt-4o-2024-08-06} (GPT-4o) \cite{openai2024gpt4o} for all segmentation and extraction calls. 
We select GPT-4o for its balance of performance, latency and cost, and because it is currently production-stable in our deployment setting.
We also assess backbone variability with \texttt{gpt-4.1-2025-04-14} (GPT-4.1) \cite{openai2025gpt41} and alternative embedders. 
GPT-4o is run with temperature~0.  
The SchemaRAG hyperparameter $k$ is fixed to 60, and all row-weights $w$ in Eq.~\ref{eq:rowembedding} are held constant (Appendix~\ref{sec:experiment}).
We repeat each main experiment three times. 

Prompts follow the structure shown in Fig.~\ref{fig:reduction}, with domain-specific tokens adjusted minimally for each dataset. 
When an input exceeds GPT-4o's 128k-token context window, we apply a truncation procedure that prunes schema rows until the prompt fits; the full algorithm is detailed in Appendix~\ref{sec:experiment}.

\subsection{Metrics}

\textbf{Accuracy.}~We report an F\textsubscript{1} score using true/false positives and false negatives defined for the structured IE task.
Multi rows are evaluated per value (“unrolled”). Invalid or out‑of‑schema outputs are discarded, and type‑coercible values are normalized (\textit{e.g.} ``4'' $\rightarrow$ 4 for a Numeric row)  before scoring. Scores are averaged over three runs.

\textbf{Latency and token use.} We report average per-transcript latency (in seconds) and average per-transcript token counts (separately for prompt/input and completion/output).
For SchemaRAG, latency includes one segmentation call plus the maximum extraction‑call duration (parallel execution).

\textbf{Significance testing.} To evaluate the statistical significance of differences in F\textsubscript{1}, latency, and token usage, the Wilcoxon signed-rank test \cite{wilcoxon1992individual} is applied to 
per-transcript results. 
We treat $p < 0.05$ as statistically significant. 
Full metrics details can be found in Appendix~\ref{sec:metrics}.

\begin{table*}
    \centering
    \caption{Average per-transcript latency and token use. Statistical significance of each condition over the previous condition is indicated by \textsuperscript{\textdagger} ($p <$ 0.001).}
    \label{tab:usage}
    \begin{tabular}{lrrrr}
    \toprule
        & \multicolumn{4}{c}{\textbf{Nursing}} \\
        \cmidrule(lr){2-5}
        & Lat. (s) $\pm$ \textsc{cv} 
        & Tok. in $\pm$ \textsc{cv} 
        & Tok. out $\pm$ \textsc{cv}
        & Total tok.
        \\
    \midrule
    Full-schema 
        & 11.3 $\pm$ 61.4\% 
        & 77,382 $\pm$ 27.8\%
        & \textbf{212} $\pm$ 134.5\%
        & 77,594
        \\ 
    SchemaRAG 
        & \textbf{6.0}\textsuperscript{\textdagger} $\pm$ 68.9\%
        & \textbf{38,707}\textsuperscript{\textdagger}  $\pm$ 137.1\%
        & 421\textsuperscript{\textdagger}  $\pm$ 142.2\%
        & \textbf{39,128}
    \\
    \textit{+ oracle} 
        & \textit{4.0}\textsuperscript{\textdagger} $\pm$ \textit{82.9}\%
        & \textit{31,721}\textsuperscript{\textdagger}  $\pm$ \textit{169.1\%}
        & \textit{415}\textsuperscript{\textdagger}  $\pm$ \textit{142.6\%}
        & \textit{32,136}
    \\

        \midrule
        & \multicolumn{4}{c}{\textbf{Amazon}} \\
        \midrule

        Full-schema 
            & \textbf{16.4} $\pm$ 27.6\%
            & \textbf{122,095} $\pm$ 2.3\%
            & \textbf{147}  $\pm$ 72.0\%
            & \textbf{122,243}
        \\
        SchemaRAG
            & 17.7 $\pm$ 35.9\%
            & 170,929\textsuperscript{\textdagger} $\pm$ 44.7\%
            & 2,786\textsuperscript{\textdagger} $\pm$ 43.2\%
            & 173,714
        \\
        \textit{+ oracle} 
            & \textit{13.0}\textsuperscript{\textdagger} $\pm$ \textit{38.7}\%
            & \textit{107,004}\textsuperscript{\textdagger} $\pm$ \textit{87.6}\%
            & \textit{3,766}\textsuperscript{\textdagger} $\pm$ \textit{51.5}\%
            & \textit{110,771}
        \\

    \bottomrule
    \end{tabular}

\end{table*}

\subsection{Baseline and oracle} 
We compare SchemaRAG on Nursing and Amazon against two baselines: i) a full-schema prompt and ii) a full-schema+segmentation prompt. 
The SchemaRAG condition represents the ``maximal'' form of our method, which always uses segmentation and uses examples, when available (\textit{i.e.}, for Nursing).
The full-schema baseline sorts rows by relevance only to include as many rows as fit into context, whereas SchemaRAG selects a small, targeted top-$k$ as its retrieval set to guide extraction. To avoid understating the baseline, we retain this relevance sort. 
Other sorting-method comparisons and additional baselines (including a lightweight table‑inspired baseline that retrieves by row name only) are provided in Appendix~\ref{sec:baselines}. We also provide an oracle upper bound for the SchemaRAG method by constructing the reduced schema directly from the reference annotated rows.

\subsection{Extraction accuracy}
Nursing F\textsubscript{1} results across all four hospital datasets are calculated as a micro-F\textsubscript{1} score. We also report macro-F\textsubscript{1} scores by averaging per-transcript F\textsubscript{1} scores across the four datasets, and use them in statistical significance testing.

Across both Nursing and Amazon, SchemaRAG outperforms all baselines in micro-F\textsubscript{1} (Table~\ref{tab:results}), with relative gains of 8.8\% on Nursing and 8.3\% on Amazon. 
These improvements persist with a GPT‑4.1 backbone and other embedders on Nursing (Appendix~\ref{sec:other_backbones}). 
On Amazon, segmentation alone improves recall (0.367 $\to$ 0.641) but degrades precision (0.661 $\to$ 0.316). When combined with SchemaRAG, we observe a recovery in precision (0.316 $\to$ 0.467), resulting in a more balanced performance profile and the highest F\textsubscript{1}.
The extraction accuracy on Amazon is consistently lower compared to Nursing, likely due to several dataset‑specific factors we discuss in Section \ref{sec:limitations}.
Finally, while oracle results significantly outperform SchemaRAG on both datasets, they establish a high theoretical ceiling for schema reduction and validate the potential for further refinements to this approach.

\subsection{Latency and token use}

\textbf{Computing latency.} In general, SchemaRAG reduces or maintains latency (Table~\ref{tab:usage}). Compared to the baseline, SchemaRAG introduces three differences
impacting latency: 
i) the added latency of segmentation;
ii) parallel extraction of $n$ segments, with total latency bounded by the longest segment's extraction time;
and iii)
the substantially reduced size of the schema used in extraction prompts.

Despite segmentation's additional latency, SchemaRAG is 47\% faster than the baseline on Nursing, owing to the far smaller number of extraction input tokens (Table~\ref{tab:usage}) and the reduction in 
longest-parallel-extraction output tokens (here reducing from 212 to 52). 
Our Amazon experiments
show slightly higher latency for SchemaRAG relative to the baseline, though the effect is not significant. SchemaRAG's $O(T)$ segmentation call (Appendix~\ref{sec:token_complexity}) dominates latency due to Amazon's larger transcripts. Thus, despite reduced schema use and extraction parallelization, SchemaRAG yields no reduction in latency on this dataset. This difference demonstrates the impact of scaling transcript size on SchemaRAG: while F\textsubscript{1}'s trajectory is unaffected by an increasing transcript length, latency may plateau or increase.

\textbf{Computing token use.} SchemaRAG reduces token cost for Nursing but raises it for Amazon (Table~\ref{tab:usage}). Although OpenAI charges 4$\times$ more for output tokens than input tokens \cite{OpenAI_pricing}, input tokens dominate token cost in our experiments, exceeding output volume by two to three orders of magnitude. 
On Nursing, SchemaRAG uses roughly twice the output tokens of the baseline, nearly reaching the oracle count, while halving input tokens.  
This suggests the baseline undergenerates, likely due to lost-in-the-middle effects. SchemaRAG's mitigation yields a 50\% reduction in total token use and 48\% in token cost.

On Amazon, SchemaRAG’s segmentation generates more parallel extraction calls due to larger average transcript sizes. 
Because each per-segment prompt includes the full transcript for context (Figure~\ref{fig:reduction}), transcript size $T$ becomes a decisive factor for token complexity, which can be modeled as $O(T^2 + T{S_\kappa}_i + TA_i)$ (Appendix \ref{sec:token_complexity}).
Consequently, SchemaRAG's input tokens exceed the baseline by about 40\%. 
SchemaRAG also yields more output tokens, approaching the oracle condition.
This suggests that the baseline also undergenerates and that SchemaRAG mitigates this issue.

Nevertheless, given SchemaRAG's parallelization of extraction and reduced schema size, we anticipate SchemaRAG could remain competitive in latency and cost for datasets with moderately sized $T$.

\subsection{Ablations} 
Since SchemaRAG is a flexible, modular method, we conduct ablations to assess the impact of different components on extraction performance. We analyze the effects of using i) supplementary examples, ii) row embeddings and example embeddings, and iii) segmentation scope. 
Guidance from labeled examples, especially when used in both retrieval and prompting, consistently improves performance. Combining row and example embeddings yields the best macro-F\textsubscript{1} (0.899 vs. 0.888 row-only and 0.798 example-only). Segmentation primarily helps retrieval (Nursing: 0.898 vs.\ 0.774 baseline; Amazon: 0.554 vs.\ 0.484 baseline). 

We also examine the effect of the hyperparameter $k$. Performance generally improves as $k$ increases until saturating, with $k{=}60$ providing a strong operating point in our experiments. 
SchemaRAG also attains consistently higher recall@k than the full-schema+segmentation baseline,  approaching the upper bound faster (\textit{e.g.}, recall@5: 0.800 vs.\ 0.749 baseline; recall@60: 0.991 vs 0.981 baseline). 
Full ablation results are in Appendix ~\ref{sec:ablation}.

\section{Conclusion}

SchemaRAG is a retrieval-augmented generation method for structured information extraction that dynamically selects a relevant subset of a complex schema based on unstructured input text. Evaluations on two real-world, large-schema datasets show that SchemaRAG improves accuracy over full-schema baselines and remains competitive in latency. 
While its impact on token cost depends on transcript size, oracle experiments show that better accuracy can be achieved while using fewer tokens. This validates the promise of more precise subschema selection as a method, positioning SchemaRAG as a practical step toward efficient and accurate LLM-based structured information extraction.

\section{Limitations}
\label{sec:limitations}
We observe consistently lower extraction accuracy on the Amazon dataset compared to the Nursing dataset. This difference likely reflects several dataset‑specific factors. The Nursing dataset uses schemas designed \textit{a priori} by hospitals to support real EHR workflows, paired with naturally occurring nurse dictations and labeled in‑domain examples. In contrast, the Amazon dataset was manually constructed from noisier source data.
Preparing the Amazon dataset for structured information extraction required substantial cleaning and transformation (Appendix \ref{sec:datasets-amazon}), and the resulting schema rows and gold labels are therefore likely imperfect. Moreover, unlike the Nursing dataset, we do not have access to additional labeled examples to assess the effect of label quality or data scale, which ablation results on Nursing suggest could be beneficial (Appendix \ref{sec:ablation-example}).
Finally, Amazon input transcripts were created by concatenating product descriptions and related metadata, resulting in longer, more diverse, and multi‑topic inputs than nurse dictations. While SchemaRAG’s segmentation strategy was held constant across datasets for consistency, it was not exhaustively optimized for this setting. These factors likely contribute to the reduced performance observed on the Amazon dataset.

\section*{Ethical Considerations}
Our clinical experiments use de‑identified, automatically transcribed nursing speech; transcription and labeling errors may propagate to extracted structures.  
We suggest a human-in-the-loop deployment model requiring clinician review and robust audit trails.
With respect to data privacy, we do not release any clinical data or prompts that could reveal protected information; e‑commerce data is derived from a publicly available sample and is used only for internal evaluation. 
\bibliography{custom}

\clearpage
\appendix
\section{Datasets}
\label{sec:datasets}

\subsection{Nursing}

\begin{figure}[h]
    \centering
    \includegraphics[width=.9\linewidth]{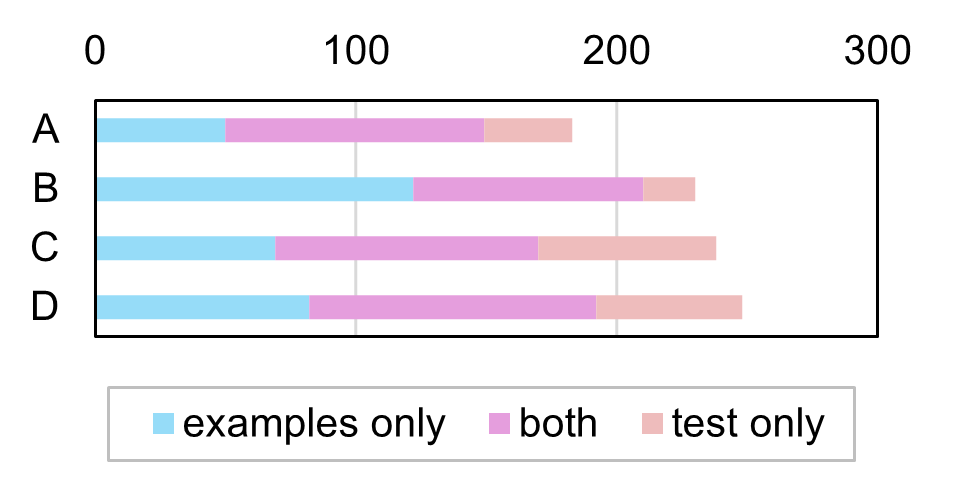}
    \captionof{figure}{Counts of unique rows that appear in each Hospital's test set, example set, or both.}
    \label{fig:coverage}
\end{figure}

Table \ref{tab:datasets} and Figure \ref{fig:coverage} 
give a breakdown of the size and content of the Nursing dataset.

\begin{table}[!h]
    \centering 
    \caption{Nursing dataset: total and unique row counts for schemas, test sets, and example sets.
    }
    \label{tab:datasets}
    \begin{tabular}{crrrrr}
        \toprule
        \textbf{Hospital} & \textbf{Sch.} & \multicolumn{2}{c}{\textbf{Test}} & \multicolumn{2}{c}{\textbf{Examples}} \\
        \cmidrule(lr){3-4}
        \cmidrule(lr){5-6}
        & & tot. & un. & tot. & un. \\
        \midrule 
        A & 1064 & 739 & 133 & 1816 & 149 \\
        B & 1118 & 308 & 108 & 881 & 210 \\
        C & 601 & 741 & 169 & 2176 & 170 \\ 
        D & 1261 & 551 & 166 & 1424 & 192 \\
        \bottomrule
    \end{tabular}
\end{table}

\subsection{Amazon}
\label{sec:datasets-amazon}


The Amazon dataset used in this work (the ``Subset'') is a subset of the publicly available sample of the Bright Data Amazon Products dataset \cite{brightdata2025} (the ``Public Sample''). While the Public Sample is clearly labeled as publicly downloadable, it does not come with a named license. Therefore, out of an abundance of caution and wish to abide by any existing copyright restrictions, we do not release our Subset derived from the Public Sample. We instead describe our process for internally creating the Subset here.

The Public Sample consists of scraped data from 1000 Amazon product webpages, including various prose features and a set of annotated key-value pairs for each product. To transform this data into a schema and test set suitable for structured information extraction, we follow the steps outlined below.

\textbf{Schema preparation.}~In the Public Sample, each product has a list of annotated groups $G = \{g_0, ..., g_n\}$ and a set of annotated key-value pairs. We form the schema $S_{\mathrm{Am}}$ by creating a distinct row for each unique $g_0$--key combination, taken from across all 1000 products. Possible values for each of these rows are collected from every instance of the $g_0$--key combination in the Public Sample. Each row's categories $C$ are the set of all groups from the products where the row's $g_0$--key combination is present.

After assembling possible values for every row in this way, closer attention is needed. We discard all rows with metadata not in English. For rows with values containing commas, semicolons, or other obviously delimited lists of choices, we split each delimited string into its own possible value and schematize the row as a Multi row. Rows that, on manual inspection, indicate numeric values or weights are schematized as Numeric or Weight rows, respectively. All remaining rows are schematized as Single rows.

Single rows for which the above process yielded only one possible option are either ``hydrated'' when reasonably completable (\textit{e.g.}, a Single row only containing a ``yes'' option is also supplied a ``no'' option) or discarded.

\textbf{Test set preparation.}~We randomly sample a maximum of two products per annotated ``department'' key, resulting in a total of 48 products in the Subset. For each of these products, we prepare a transcript by concatenating the product title, description, top review, product dimensions, weight, manufacturer, and bullet point prose features into a single string. 

We also manually review the annotated key-value pairs (excluding ``department'') to ensure they can be substantiated by the transcript content. Key-value pairs that cannot be reasonably inferred from the transcript, such as serial numbers or model names that are never mentioned, are discarded from the annotations. When an option from the schema is clearly available and applicable to the transcript, we add it to the annotations. For example, if the transcript mentions a product is ``red and blue,'' and there exist ``red'' and ``blue'' Multi options in the appropriate schema row, but the original annotated value was ``red'' alone, we add ``blue'' to the values for that annotated key-value pair. Annotations that are plainly inconsistent with the transcript are corrected. For example, if the transcript mentions a quantity of ``3'' and the original annotated value was ``2,'' we correct the annotation to ``3.'' We also remove any duplicate key-value pairs from the annotations.

\section{Schema representation}
\label{sec:schema_representation}


An example of a reduced schema $S_\kappa $ that SchemaRAG might produce given the transcript $T$ \textit{``pain is at 5 out of 10''}:

\begin{lstlisting}[
    basicstyle=\tiny,
    language=json,
    firstnumber=1,
    columns=fullflexible,
    label={lst:reduced_schema}
]
{
  "reduced_rows": [
    {
      "id": "0003",
      "type": "Number",
      "row": "Pain Level",
      "category": ["Pain Test"]
    },
    {
      "id": "0001",
      "type": "Single",
      "row": "In pain now",
      "category": ["Pain Test"],
      "values": ["Yes", "No"]
    }
 
}
\end{lstlisting}
A corresponding example of that reduced schema in JSON Schema format, which we present to the LLM for filling:
\begin{lstlisting}[
    basicstyle=\tiny,
    language=json,
    firstnumber=1,
    columns=fullflexible,
    label={lst:reduced_schema}
]
{
  "type": "object",
  "properties": {
    "Pain Test": {
      "type": "object",
      "properties": {
        "Pain Level": {
          "type": "number"
        },
        "In pain now": {
          "type": "string",
          "enum": ["Yes", "No"]
        }
      }
    }
  }
}
\end{lstlisting}
\section{Models and hyperparameters}
\label{sec:experiment}

\subsection{Row weights}
All weights $w$ in Eq. \ref{eq:rowembedding} are fixed throughout and were tuned by grid search on preliminary proprietary held-out test sets drawn from Hospitals A--D; the values of $w$ and metadata to which they apply are proprietary, but include weighting a row's name, category(ies), and possible value(s).

\subsection{Prompt truncation method}
To ensure that the prompt length stays within the input token limit, we apply a truncation strategy that dynamically adjusts the number of schema rows to be included in the prompt in the full schema experiments (with no examples). The procedure is as follows:

\begin{enumerate}
\item First, construct the full prompt using all schema rows and compute the total token count, including the instruction prompt and the schema (in JSON Schema format as in Table~\ref{tab:exampleschema}). If the token count is already within the limit, return the prompt as-is.

\item Otherwise, compute a relevance score $\rho$ for each row (\textit{i.e.}, cosine similarity to the input), and sort the rows in descending order of relevance so that the least relevant rows can be removed from the end of the list during truncation. 

\item Given the sorted list of rows, initialize two pointers:  \texttt{left} at 0 and  \texttt{right} at the total number of rows in the schema.

\item While \texttt{left} $\leq$ \texttt{right}, compute the midpoint index \texttt{mid}.

\item Construct a temporary prompt using only the first \texttt{mid} rows and compute the total token count, including the instruction prompt and schema (in JSON Schema format as in Table~\ref{tab:exampleschema}).

\item If the total token count is within the limit, update \texttt{left = mid + 1} to try including more rows.

\item If the token count exceeds the limit, update \texttt{right = mid - 1} to reduce the number of rows. 

\item Repeat until the maximum valid number of matches is found. The final prompt is then constructed using this schema subset.
\end{enumerate}
\section{Metrics}
\label{sec:metrics}

\subsection{Accuracy computation}
We compute F\textsubscript{1} from true positives (TP), false positives (FP), and false negatives (FN) defined at the row–value level. 
For Multi rows, each predicted value is treated independently (“unrolled”) to avoid crediting extra unattested values. 
LLM responses that cannot be mapped to any schema row or valid value type are marked invalid and excluded from scoring. 
When a predicted value is type‑coercible (\textit{e.g.}, the string ``4'' for a Numeric row), we normalize it before comparison. 
All F\textsubscript{1} results are averaged over three independent runs, with coefficients of variation computed across runs.

\subsection{Latency and token accounting}
We measure per‑transcript latency and token usage separately for (i) prompt/input tokens and (ii) completion/output tokens.
For the baseline (no segmentation), latency is simply the extraction call duration.
For SchemaRAG, we measure (a) the segmentation call and (b) the $n$ extraction calls; because extraction calls run in parallel, the effective extraction latency is the maximum of these $n$ durations.
Final reported latency and token counts are averaged over three runs for each experimental condition, also with coefficients of variation.

\subsection{Significance testing}
To assess whether differences in F\textsubscript{1}, latency, or token usage are statistically reliable, we apply the Wilcoxon signed-rank test \cite{wilcoxon1992individual} to per-transcript metrics, averaged over three runs. 
This non-parametric test is chosen because Shapiro–Wilk tests \cite{shapiro1965analysis} indicate that per-transcript distributions significantly deviate from normality, often exhibiting skewness or heavy tails.
Unless otherwise stated, we treat $p < 0.05$ as statistically significant.

\section{Baselines}
\label{sec:baselines}

\subsection{Baseline sorting controls and results}
The baseline used in this work sorts the rows by relevance before it estimates how many rows are able to fit in the prompt, then truncates those that do not fit. Unlike SchemaRAG, which uses a small, targeted $k$, the baseline aims to completely fill the prompt up with as many rows from the full schema as possible, stopping only when the context limit is reached.

The relevance sort is a pragmatic measure that is especially necessary for the Amazon dataset, whose schema is very large. (An average of 86\% of its rows must be truncated for final extraction prompts to fit in the context.) If we were to truncate this schema without sorting on relevance, the chances of any correct row surviving truncation is low, because the schema comes by default in an arbitrary order. This would lead to artificially low performance and would not be an interesting condition. To ensure consistency in method, the same sorting approach for Nursing is applied, even though its full schemas fit in context.

To compare the relevance sorting approach with other approaches, we ran additional full-schema experiments on Nursing. As shown in Table \ref{tab:sorting_methods}, the relevance-based sorting approach provides a more competitive baseline. Notably, SchemaRAG's relative gains would appear even more significant if compared against a baseline lacking this sorting method.

\begin{table}[t]
    \centering
    \caption{Comparison of schema sorting methods on macro-F\textsubscript{1} performance (run 3$\times$).}
    \label{tab:sorting_methods}
    \begin{tabular}{lr}
        \toprule
        \textbf{Sorting Method} & \textbf{macro-F\textsubscript{1}} \\
        \midrule
        As-is, arbitrary order & 0.762 \\
        \makecell[lt]{By JSON keys \\ \quad (alphabetical on hierarchy)} 
            & 0.772 \\
        \makecell[lt]{By relevance \\ \quad (proposed method)} 
            & \textbf{0.851*} \\
        \bottomrule
        \multicolumn{2}{l}{
            \small * Statistically significant ($p < 0.001$) vs. other methods.
        }
    \end{tabular}
\end{table}

\subsection{Additional baseline}
We report an additional baseline here that is inspired from tabular, rather than text-based, RAG strategies. This method uses only each schema row’s embedded name for retrieval, which is a lightweight grounding strategy somewhat analogous to TableRAG’s ``ReadSchema'' approach \cite{chen2024tablerag} and minimally exploits the structured nature of the schema by using only one element of its metadata. We report micro-F\textsubscript{1} (averaged over three runs) on the Nursing and Amazon datasets (Table~\ref{tab:additional_baseline}).

\begin{table}[ht]
    \centering
    \caption{Comparison of additional baselines.}
    \label{tab:additional_baseline}
    \begin{tabular}{lcc}
        \toprule
        \textbf{Method} & \textbf{Nursing} & \textbf{Amazon} \\
        \midrule
        Full-schema & 0.844 & 0.471 \\ 
        Row-name-only & 0.800 & 0.341 \\ 
        SchemaRAG & \textbf{0.918} & \textbf{0.510} \\ 
        \bottomrule
    \end{tabular}
\end{table}

SchemaRAG outperforms this lightweight table-inspired baseline as well as the full-schema, information-extraction-inspired baseline.
\section{Other backbone models}
\label{sec:other_backbones}

SchemaRAG is fully modular and requires no fine-tuning, so applying it to other LLMs is straightforward. To assess backbone variability, we replicate the Nursing experiments with \texttt{gpt-4.1-2025-04-14} (GPT-4.1) and report macro-F\textsubscript{1} (averaged over three runs). Results show that even with a more powerful and capable LLM, SchemaRAG significantly outperforms the full-schema baseline (Table~\ref{tab:gpt41_macro}). Even as context windows expand in state-of-the-art models, SchemaRAG offers continued value by steering model attention toward the most relevant schema elements. This improves extraction accuracy and mitigates "lost-in-the-middle" effects that can still arise even with larger context capacities.

\begin{table}[ht]
\centering
\caption{Nursing macro-F\textsubscript{1} (3 runs averaged) with GPT-4o vs.\ GPT-4.1. Asterisks indicate significance over the corresponding full-schema baseline ($p<0.005$).}
\label{tab:gpt41_macro}
\begin{tabular}{lcc}
\toprule
\textbf{Method} & \textbf{GPT-4o} & \textbf{GPT-4.1} \\
\midrule
Full-schema   & 0.851 & 0.868 \\
SchemaRAG     & \textbf{0.899}$^{*}$ & \textbf{0.900}$^{*}$ \\
\bottomrule
\end{tabular}
\end{table}

Because retrieval relies on semantic similarity, the choice of embedder can influence performance. We run a single-pass (x1) check on the Nursing dataset to assess sensitivity to different off‑the‑shelf embeddings; results are reported as micro‑F\textsubscript{1}. Across embedders, SchemaRAG consistently outperforms the full‑schema baseline, indicating that while the encoder choice can shift absolute accuracy, the method’s relative gains are robust (Table~\ref{tab:embedder_swap_table}).

\begin{table}[t]
\centering
\caption{Nursing micro-F\textsubscript{1} (single run) across \texttt{text-embedding-*} embedders.}
\label{tab:embedder_swap_table}
\begin{tabular}{lcc}
\toprule
\textbf{Embedder} & \textbf{Full-schema} & \textbf{SchemaRAG} \\
\midrule
\texttt{-ada-002} & 0.847 & \textbf{0.915} \\
\texttt{-3-small}            & 0.853 & \textbf{0.902} \\
\texttt{-3-large}            & 0.849 & \textbf{0.904} \\
\bottomrule
\end{tabular}
\end{table}

\section{Token complexity}
\label{sec:token_complexity}


We model the token complexity of SchemaRAG in this way.

For all API LLM-driven methods, the worst-case upper bound on the number of tokens used in a single LLM call (prompt and completion) is the token limit imposed by the service, as the LLM may or may not follow instructions to respond in the form of the provided schema and could exceed it. In this analysis, we assume a typical case to involve a well-behaved LLM that follows instructions to respond in the desired format. 

SchemaRAG consists of two steps: segmentation and extraction. 
For this analysis, we assume that the transcript text $T$ to be segmented contains non-duplicated, non-contradictory extractable facts that are well-distributed throughout the text, so that a well-behaved LLM segmenter will yield segments of roughly equal length.
In segmentation, the input prompt contains the full transcript, yielding complexity $O(T)$, and the segmented output is expected to be of roughly equal size, also $O(T)$. Input and output together yield a complexity of $O(T)$ (with the doubling of the single term from input and output an ignorable constant).

In extraction, we assume that the hyperparameter $k$ is set such that it will exhaust neither the number of available schema rows nor the number of examples in a full example set $A$. Each segment $t_i$'s extraction prompt contains a number of dynamically-sized items: $t_i$, the full transcript $T$ (provided for context),
the reduced schema ${S_\kappa}_i$, and any retrieved examples $A_i$. We can model this complexity as $O(t_i + T + {S_\kappa}_i + A_i)$, which reduces to $O(T + {S_\kappa}_i + A_i)$ (because $t_i$ is related to $T$ by an ignorable constant).

Segmentation yields $n$ segments, each with an extraction prompt.
Given our expectation of segments of roughly equal length and roughly equal extractable fact density, we can also suppose that SchemaRAG would yield $n$ reduced schemas with a roughly equal number of rows, and $n$ retrieved example sets with a roughly equal number of examples, also assuming that the available full available example set $A$ uniformly covers the test set rows. 

On these assumptions, we can model all $n$ input extractions as having complexity 
$O(nT + n{S_\kappa}_i + nA_i)$.
Since $n$ is a linear factor of $T$, 
this can be rewritten as $O(T^2 + T{S_\kappa}_i + TA_i)$.

Extraction output complexity is $O(n{{S_\kappa}_i})$, since in the worst case, the entire reduced schema could be filled in each segment's extraction. This can be rewritten as $O(T{S_\kappa}_i)$.

Total extraction complexity for a transcript, across all segments, is therefore 
$O(T^2 + T{S_\kappa}_i + TA_i)$
(as the doubling of $O(T{{S_\kappa}_i})$ in input and output complexity is an ignorable constant).

Finally, total complexity of SchemaRAG including segmentation and extraction is 
$O(T^2 + T{S_\kappa}_i + TA_i + T)$,
which can be simplified to 
$O(T^2 + T{S_\kappa}_i + TA_i)$
(because $T \leq T^2$).

\section{Ablation studies}
\label{sec:ablation}

\subsection{Effect of example method} 
\label{sec:ablation-example}
\begin{figure}[t]
    \centering 
    \includegraphics[width=\linewidth]{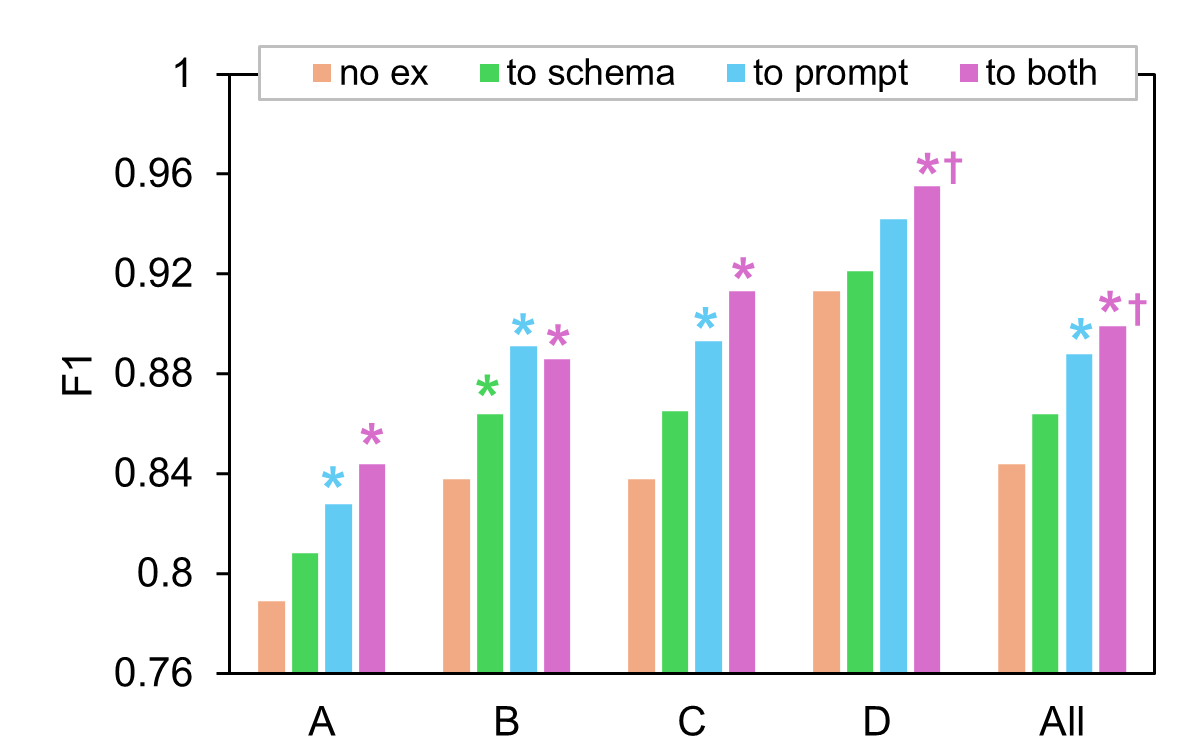}
    \caption{Effect of example use method on SchemaRAG F\textsubscript{1} on the Nursing dataset. * indicates significance over the no-example baseline, and \textsuperscript{\textdagger} over the ``to-prompt'' method, at $p <$ 0.01.}
    \label{fig:exampleablation}
\end{figure}

An advantage of SchemaRAG is its openness to the use of any extra annotated data. When a set of annotated examples $A$ is available, it could be incorporated in three ways: i) as example embeddings that drive the schema reduction process (``to-schema''), ii) as in-context learning material that is injected into the prompt (``to-prompt''), or iii) both (``to-both''). We examine the effect of these example methods on the extraction performance of SchemaRAG on the Nursing dataset, breaking down by hospital (Figure \ref{fig:exampleablation}).

Across all four hospitals, the ``to-both'' method significantly improves extraction accuracy over the no-example baseline. This demonstrates that incorporating guidance from examples at both the retrieval and inference stages is effective across different schema structures.

The gain from the ``to-schema'' method alone is not statistically significant. In contrast, the ``to-prompt'' method yields a statistically significant improvement alone, indicating that few-shot supervision at inference time plays a central role in guiding accurate extraction. The performance of the ``to-both'' method also significantly exceeds that of the ``to-prompt'' method.
This suggests that inclusion of example embeddings in retrieval, in addition to in-prompt example text,
yields complementary information that helps improve alignment between the reduced schema and relevant prompt content.

\subsection{Effect of row and example embeddings} 
We examine how row embeddings and example embeddings, used independently and together during the retrieval stage, affect extraction performance on the Nursing dataset. 
All pairwise differences between conditions are statistically significant (Table~\ref{tab:embeddingablation}).

Using only example embeddings yields the lowest performance. This can be partly attributed to the fact that each per-hospital example set does not cover all rows present in the per-hospital test set. Some test rows have no corresponding examples (see Figure~\ref{fig:coverage}), placing a fundamental limit on row recall. Additionally, the example set contains schema rows that are not part of the test set, which introduces noise. By
contrast, row-level embeddings provide complete coverage and ensure that all schema entries are potentially retrievable. 

The superior performance of combining both embeddings highlights their complementary nature: row embeddings offer broad coverage, while example embeddings provide context-rich signals when available.

\subsection{Effect of segmentation} 
We evaluate the role of segmentation in SchemaRAG by comparing three conditions: i) no segmentation at all, ii) segmentation used only for scoping the RAG process (with the full transcript and the union over all segments' resulting reduced schemas as the $S_\kappa$  provided to a single LLM call at extraction), iii) and full segmentation used both in RAG and prompt construction over per-segment extraction calls (see Table~\ref{tab:segablation}). 

On the Nursing dataset, both segmentation conditions significantly outperform the no-segmentation baseline. On the Amazon dataset, segmentation for scoping RAG significantly outperforms no segmentation, though its advantage over full segmentation used both in RAG and prompt construction is not statistically significant. These results suggest that segmentation benefits SchemaRAG primarily by guiding schema selection during the retrieval phase. Additional gains from segment-level extraction itself
depend on the domain and dataset characteristics.

\begin{table}
    \centering
    \caption{Effect of row and example embedding availability in SchemaRAG retrieval ($k=60$) on average per-transcript extraction performance on the Nursing dataset. All pairwise differences are statistically significant, indicated by \textsuperscript{\textdagger} ($p < 0.001$).}
    \label{tab:embeddingablation}
    \begin{tabular}{lll}
        \toprule
        \textbf{Row} & \textbf{Example} & \textbf{macro-F\textsubscript{1}} \\
        \midrule 
        yes & no & 0.888\textsuperscript{\textdagger} \\
        no & yes & 0.798\textsuperscript{\textdagger} \\
        yes & yes & \textbf{0.899}\textsuperscript{\textdagger} \\
        \bottomrule
    \end{tabular}
\end{table}

\begin{table}
    \centering
    \caption{Effect of segmentation scope on extraction performance of SchemaRAG. Statistical significance vs. no-segmentation SchemaRAG baseline is indicated by 
    * ($p <$ 0.05) and
    \textsuperscript{\textdagger} ($p < 0.001$).}
    \label{tab:segablation}
    \begin{tabular}{lll}
        \toprule
        \textbf{Segmentation scope} & \multicolumn{2}{c}{\textbf{macro-F\textsubscript{1}}} \\
        & Nursing & Amazon \\
        \midrule 
        Baseline & 0.774 & 0.484\\
        \midrule 
        RAG-only & 0.898\textsuperscript{\textdagger} & \textbf{0.554}* \\
        RAG + prompt & \textbf{0.899}\textsuperscript{\textdagger} & 0.515 \\
        \bottomrule
    \end{tabular}
\end{table}

\subsection{Effect of \boldmath$k$} 
\begin{figure}[!t]
    \centering
    \includegraphics[width=\linewidth]{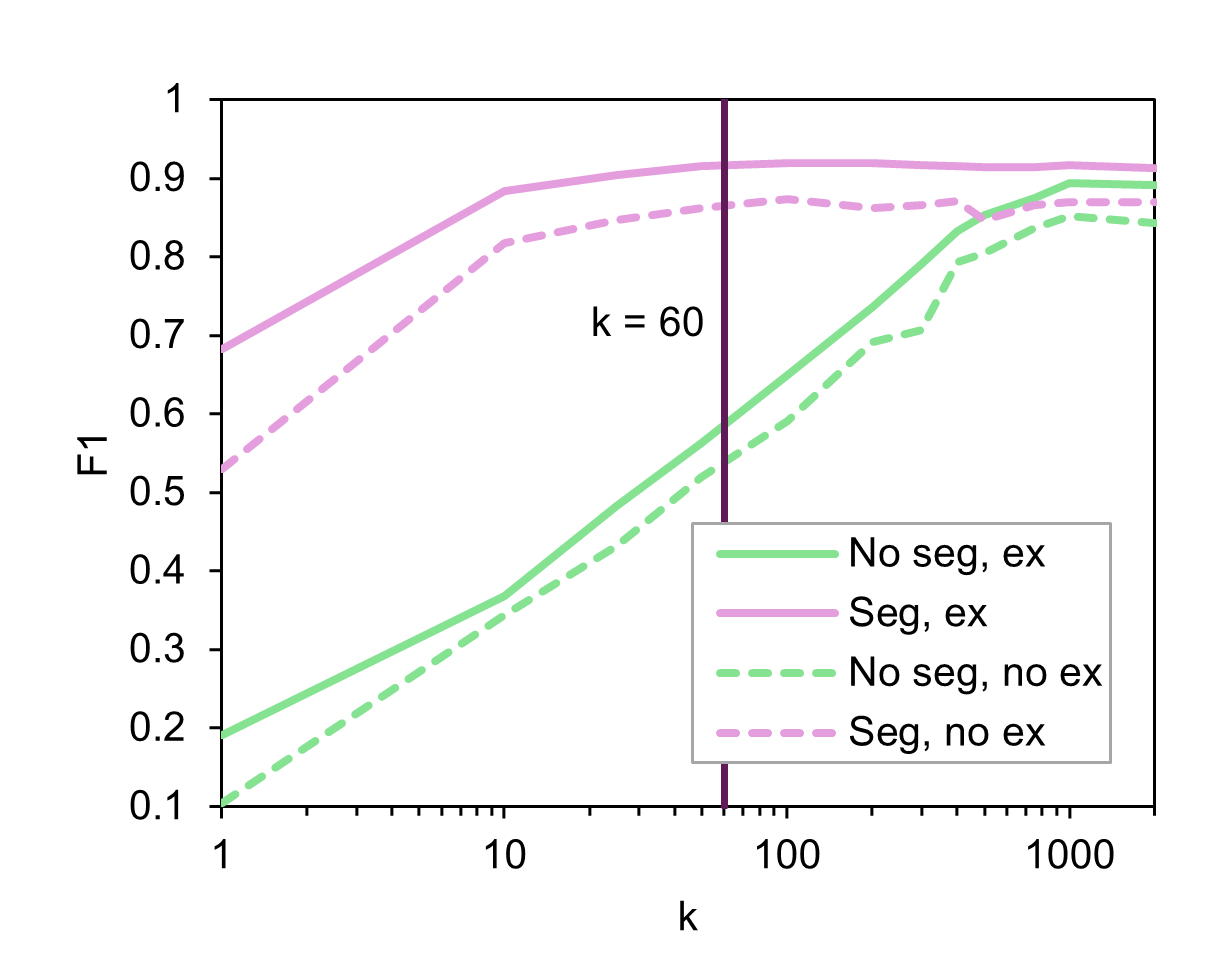}
    \caption{Effect of $k$ on SchemaRAG F\textsubscript{1} on Nursing.}
    \label{fig:varyk}
\end{figure}

We examine the effect of the hyperparameter $k$ on the extraction performance of SchemaRAG on the Nursing dataset. $k$ is the number of embeddings consulted during the schema reduction process and roughly correlates with $\kappa$, the final number of rows in the reduced schema. We vary $k$ to produce reduced schemas of size $\kappa$ between 1 and 1100 with and without segmentation, and with and without examples, running each condition only once (Figure \ref{fig:varyk}). 

In general, increasing $k$ improves performance to a point, as the relevant rows appear in the reduced schema; but performance saturates and declines as $k$ increases, as the reduced schema becomes too large and the LLM is unable to focus on the relevant rows. This effect holds over all segmentation and example conditions, though using segmentation and examples shift the optimal $k$ to a lower value.

\begin{table}[t]
    \centering
    \caption{Comparison of Precision and Recall between the Full-schema baseline and SchemaRAG at various levels (@$k$).}
    \label{tab:results_comparison}
    \begin{tabular}{@{}lccccc@{}}
        \toprule
        \textbf{Method} & \textbf{@\boldmath$k$} & \textbf{Precision} & \textbf{Recall} \\ \midrule
        Full-schema + seg. & 1  & 0.724 & 0.431 \\
                                   & 5  & 0.275 & 0.749 \\
                                   & 10 & 0.161 & 0.857 \\
                                   & 30 & 0.060 & 0.948 \\
                                   & 60 & 0.031 & 0.981 \\ 
        \midrule
        \textbf{SchemaRAG}         & 1  & 0.837 & 0.503 \\
                                   & 5  & 0.288 & 0.800 \\
                                   & 10 & 0.163 & 0.886 \\
                                   & 30 & 0.060 & 0.959 \\
                                   & 60 & 0.031 & 0.991 \\
        \bottomrule
    \end{tabular}
\end{table}

\subsection{Retrieval upper bound via Precision/ Recall@\boldmath$k$} 
We compare SchemaRAG to a full-schema baseline that uses segmentation to eliminate the effect of segmentation on top-$k$ counting. Each segment of a transcript yields its own SchemaRAG-retrieved set of rows; we take the top $k$ of each segment, then combine these and compare to the gold-standard annotations for the entire transcript.

SchemaRAG achieves consistently higher recall@$k$ than the full-schema+segmentation baseline at equivalent k values (Table~\ref{tab:results_comparison}). This suggests that SchemaRAG yields more of the annotated rows than the naïve baseline and approaches the upper bound of retrieval faster with increasing $k$.

Similarly, SchemaRAG exhibits higher precision@$k$ at low $k$  (Table~\ref{tab:results_comparison}), suggesting that the annotated rows it finds are more densely packed near the top of the retrieval. (At high $k$, the baseline and SchemaRAG have similar precisions because there are more cases of fewer than $k$ annotated rows over each transcript).

\end{document}